# Electric field induced semiconductor-to-metal phase transition in vertical $MoTe_2$ and $Mo_{1-x}W_xTe_2$ devices

*Feng Zhang[1,2], Sergiy Krylyuk[4,5], Huairuo Zhang[4,5*], Cory A. Milligan[1,3], Dmitry Y. Zemlyanov[1], Leonid A. Bendersky[5], Albert V. Davydov[5] and Joerg Appenzeller[1,2*]*

[1]Birck Nanotechnology Center, Purdue University, West Lafayette, Indiana 47907, USA.
[2]Department of Electrical and Computer Engineering, Purdue University, West Lafayette, Indiana 47907, USA.
[3]School of Chemical Engineering, Purdue University, West Lafayette, Indiana 47907, USA.
[4]Theiss Research, Inc., La Jolla, California 92037, USA.
[5]Materials Science and Engineering Division, National Institute of Standards and Technology (NIST), Gaithersburg, Maryland 20899, USA.

[*]**Corresponding Authors:** appenzeller@purdue.edu (J.A.); huairuo.zhang@nist.gov (H.Z.)

**Abstract**

Over the past years, transition metal dichalcogenides (TMDs) have attracted attention as potential building blocks for various electronic applications due to their atomically thin nature. An exciting development is the recent success in "engineering" crystal phases of TMD compounds during the growth due to their polymorphic character. Here, we report an electric field induced reversible engineered phase transition in vertical 2H-$MoTe_2$ devices, a crucial *experimental* finding that enables electrical phase switching for these ultra-thin layered materials. Scanning tunneling microscopy (STM) was utilized to analyze the TMD crystalline structure after applying an electric field, and scanning tunneling spectroscopy (STS) was employed to map a semiconductor-to-metal phase transition on the nanoscale. In addition, direct confirmation of a phase transition from 2H semiconductor to a distorted 2H′ metallic phase was obtained by scanning transmission electron microscopy (STEM). $MoTe_2$ and $Mo_{1-x}W_xTe_2$ alloy based vertical resistive random access memory (RRAM) cells were fabricated to demonstrate clear reproducible and controlled switching with programming voltages that are tunable by the layer thickness and that show a distinctly different trend for the binary compound if compared to the ternary materials.

**Main text**

Many applications such as memristors[1], micro-motors[2], electronic oscillators[3], and sensors[4] greatly benefit from recent trends in the area of "phase engineering". The most prominent materials that are explored in this context are $VO_2$ and $NbO_2$ which can both undergo a Mott metal-to-insulator transition[5] and amorphous-to-crystalline phase change materials as $Ge_2Sb_2Te_5$[6]. Recently, transition metal dichalcogenides (TMDs) attracted considerable attention in the field of 2D phase engineering due to their polymorphic character. TMDs exist in various crystalline phases which exhibit semiconducting, semimetallic and metallic properties. In particular, experimental phase diagram data and density functional theory (DFT) calculations indicate that the most thermodynamically stable phase at room-temperature for molybdenum and tungsten containing dichalcogenides is the semiconducting hexagonal (2H) phase, with the exception of $WTe_2$, for which the metallic octahedral ($T_d$) crystal structure is stable at standard conditions. However, since for some TMD compounds the energetic difference between the various phases is rather moderate[7,8], several groups are working on phase engineering of TMDs. For example, Py et al.[9] demonstrated a 2H to 1T phase transition in $MoS_2$ through the intercalation of lithium and Liu et al.[10] introduced an in-situ phase transition in $MoS_2$ by means of electron beam irradiation. However, $MoS_2$ might not be the ideal candidate for TMD phase engineering. Due to a very low energy difference between the metallic (1T′) and the semiconducting (2H) phase of $MoTe_2$, it is considered the most promising TMD material for phase engineered applications[7,8]. Experimental results on $MoTe_2$ include strain-induced semiconductor-to-metal transitions in $MoTe_2$ at room temperature[11] and growth controlled transformations between the 1T′ and 2H phase by means of tellurization rate[12] and temperature control[13,14]. Moreover, Empante et al.[15] recently demonstrated three distinct structural phases (2H, 1T and 1T′) in $MoTe_2$ when employing chemical vapor deposition (CVD) with controlled quenching. However, device compatible methods to switch between the different phases in $MoTe_2$ have yet to be reported.

Ultimately, for device applications, controlling the phase of a TMD by means of an electric field and introducing a semiconductor-to-metal phase transition in this way is most desirable. Here we demonstrate *experimentally* an electric field induced reversible semiconductor-to-metal phase transition in vertical 2H-$MoTe_2$ and 2H-$Mo_{1-x}W_xTe_2$ devices. Scanning tunneling microscopy (STM) was carried out to evaluate the structural impact of the electric field and scanning tunneling spectroscopy (STS) was utilized to unambiguously demonstrate the localized semiconductor-to-metal phase transition. Scanning transmission electron microscopy (STEM) showed that a distorted metallic 2H′ phase formed in 2H-$MoTe_2$ and 2H-$Mo_{1-x}W_xTe_2$ based RRAM devices that consist of a sandwich structure of TMD layers between metal electrodes after electric field application. RRAM cells with areas in the 0.1 $\mu m^2$ range displayed set voltages that are approximately linearly proportional to the flake thickness.

### 2H-$MoTe_2$ and 2H-$Mo_{1-x}W_xTe_2$ based RRAM

Fig. 1(a) shows a schematic as well as optical and SEM images of a typical vertical TMD RRAM device under investigation. Top contact areas are approximately 0.1 $\mu m^2$. Our device design ensures that only vertical transport occurs from one to the other electrode without any lateral transport contributions. Because of the large aspect ratio between the top contact area and the flake thickness, spreading resistance contributions can be ignored and the active device area is identical to the top contact area. Area normalized I-V curves of exemplary vertical $MoTe_2$, $WSe_2$ and $MoS_2$ devices are shown in Fig. 1(b). For all measurements in this article the bottom electrode was grounded. Experimental current densities follow the expected trend with thickness, i.e. higher current densities are observed for thinner flakes. The trend of higher current levels for

MoS$_2$ if compared to MoTe$_2$ and WSe$_2$ is in qualitative agreement with our findings on the height of Schottky barriers (SBs) in lateral devices from the same materials[16]. Device characteristics are reproducible and do not change substantially after multiple scans between -1 V and 1 V.

The situation however changes when the voltage range is extended. Vertical TMD devices from MoTe$_2$ and WSe$_2$ can transition into a low resistive state (LRS) as illustrated in Fig. 1(c) and (d) at a set voltage (here $V_{SET}$ = 2.3 V for MoTe$_2$) that depends on the flake thickness. The details on the forming process are provided in the Supplementary Material, where Fig. S1 (a) and (b) show characteristic I-V curves for 10 nm and 15 nm thick MoTe$_2$ flakes, respectively. After the forming event, device characteristics can be cycled to exhibit typical bipolar RRAM type of behavior in terms of: a) remaining in their LRS when no voltage is applied, b) preserving the low resistive state over an appreciable voltage range until a sufficient reset voltage (here $V_{RESET}$ = -1.5 V) of polarity opposite to the set voltage is reached, and c) remaining in their respective high resistive state (HRS) until the set voltage is reached. Note that after the forming has occurred (black symbols in Fig. 1(c)) the HRS always remains more conductive than the original state of the device (red symbol in Fig. 1(c)) indicating that a permanent electronic change has occurred in the device. For the case displayed in Fig. 1(c), the current ratio between the HRS and the LRS is about 50 when the compliance is set to 400 μA.

Compared to the behavior of MoTe$_2$, the RRAM effect in WSe$_2$ devices is much smaller. A stable LRS and HRS is achieved after forming, where the switching characteristics (see red and blue arrows in Fig. 1(d)) are consistent with the expected RRAM behavior and inconsistent with simple hysteresis effects. However, switching between the HRS and LRS is not abrupt, making WSe$_2$ a poor candidate for RRAM applications.

For vertical MoS$_2$ devices we never observed any RRAM effect for "clean devices". Note that any fabrication of RRAM devices as described here holds the risk of contamination of the bottom electrode-to-TMD and TMD-to-top electrode interface. In a number of instances when device characteristics did not show the expected current levels, e.g., in case of MoS$_2$ gave rise to current densities well below $10^4$ A/cm$^2$ at 1 V before forming for flake thicknesses below 25 nm, some changes in the device resistance can be observed that may be interpreted as RRAM behavior. However, these effects are not reproducible among devices and are likely related to the forming of filaments in an unintentionally formed oxide layer at the electrode/TMD interface. Once the intrinsic vertical current levels over $10^4$ A/cm$^2$ were reached in MoS$_2$, no RRAM behavior was observed in those devices.

In order to further explore the switching mechanism in TMDs, exfoliated MoTe$_2$ based RRAM cells with thicknesses between 6 nm and 36 nm were fabricated. All cells were nonvolatile and stable. As an example, Fig. S2(b) shows read disturb measurements on a representative MoTe$_2$ device. Before the forming process, the current per unit area through the vertical structures scales approximately with the flake thickness (see Fig. 1(b)). However, once the system transitions into its LRS, current levels (below the compliance) occur to be rather similar and do not show any coherent scaling trend with the flake thickness or active device area. This statement is consistent with the notion that the formation of conductive filaments that enable the LRS is not uniform but gives rise to local current paths. For flake thicknesses from 6 nm to 36 nm, the set voltages can be tuned from 0.9 V to 2.3 V (see Fig. 2(a) and (c)), corresponding to moderate fields in the 1MV/cm range. Noteworthy, the RRAM behavior was independent of the contact metal used. For example, employing Ni instead of Ti/Ni as top electrode resulted in the same RRAM performance as reported here.

Next, we explore the impact of the material preparation and composition on RRAM characteristics by extending experiments to 2H-$MoTe_2$ obtained using different approaches and further to 2H-$Mo_{1-x}W_xTe_2$ alloys. Similar switching characteristics of $MoTe_2$ devices fabricated using either commercial material or crystals synthesized in this work using different temperatures and transport agents (see Materials and Methods) indicate that the observed RRAM effect is not related to the processing conditions (Fig. 2(c)). $Mo_{1-x}W_xTe_2$ devices exhibit very similar switching behavior and their characteristics also depend monotonically on the flake thickness (see Figs. 2(b) and (c)). However, while we are currently unable to resolve a quantitative trend of the set voltages as a function of W-content, the set voltages for the $Mo_{1-x}W_xTe_2$ alloys show a tendency of being smaller than for the $MoTe_2$ devices (see Fig. 2(c)), implying that the critical electric field needed to trigger the RRAM behavior may have been reduced in alloys. DFT calculations[8,17,18] and recent experimental results[19] suggest that substitution of Mo by W in $MoTe_2$ and creation of a $Mo_{1-x}W_xTe_2$ alloys reduces the energy difference between 2H and 1T' phases, which in turn should reduce the set voltages of $Mo_{1-x}W_xTe_2$ RRAM devices as compared with their $MoTe_2$ counterparts. We believe that Fig. 2(c) shows a first experimental evidence of this trend.

**Electric field induced semiconductor-to-metal transition**
In general, the electroforming process in RRAM devices aims at creating a conductive filament by applying a sufficiently high electrical bias, which in turn results in an electric field and Joule heating inside the sample. Both, the field and the heating can result in the formation of conductive filaments that define the LRS while the HRS is characterized by the absence of these filaments[20]. In order to confirm the formation of conductive filaments in the case of $MoTe_2$ RRAM cells, conductive AFM (C-AFM) measurements were carried out. First, a fully functional $MoTe_2$ device was biased to form the LRS, and then wet chemical etching as described in the Supplemental Information section S9 was used to remove the top electrode. This approach allows access to the TMD surface after the presumed filament formation to perform a local analysis of the surface resistivity after the forming process had occurred. As shown in inset of Fig. 3(a) by the arrow, a bright spot of ~ 80 nm in diameter, which was formerly covered by the top electrode (red marked rectangle), is indicative of a current path through the TMD. For comparison, those $MoTe_2$ flakes that did not undergo a forming process show a uniform, highly resistive surface (see Fig. 3(b)).

In order to explore the filament formation mechanism in case of MoTe$_2$, STM was used to perform a detailed surface analysis at room temperature. Fig. 3(c) shows a representative STM image of the pristine exfoliated MoTe$_2$ flake. The atomic surface structure of the 2H-phase of the MoTe$_2$ single crystal demonstrates the expected C$_3$ symmetry with an interatomic distance of 0.34 nm. Imaging was performed with a bias voltage of -0.9 V. To mimic the situation in MoTe$_2$ RRAM cells under forming conditions, a bias of -3 V was applied to a contact underneath the TMD relative to the STM tip while scanning. Next, an STM image was taken again at a bias of -0.9 V. As apparent from Fig. 3(d) the surface image appears drastically different after voltage application, and non-uniform bright regions are clearly visible in the figure. A corrugation profile measured along the bright region is shown in Fig. S4(c). 'Protrusions' with a height of 0.3 nm to 0.6 nm on the sample surface are clearly observed, which are interpreted as possible topographic changes in conjunction with changes in the local density of states (LDOS) (see discussion below). Next, scanning tunneling spectroscopy (STS) measurements were performed at four distinct locations as marked by the colored squares 1 through 4 in Fig. 3(d). Locations 1 and 2 fall into the region of alteration while 3 and 4 are located in the unperturbed region of the sample that we will label as "pristine" in the following discussion. Fig. 3(e) shows the obtained I-V characteristics at the four locations. Higher current levels in particular for small applied voltages are observed for locations 1 and 2 (the modified areas) if compared with 3 and 4 (pristine areas). The corresponding dI/dV curves shown in Fig. 3(f) are a measure of the local density of states (LDOS). While characteristics obtained for the pristine areas (locations 3 and 4) indicate the presence of a bandgap in the range of 0.85 eV, consistent with the extracted bandgap of 2H-MoTe$_2$ from electrical measurements[16], the LDOS of locations 1 and 2 shows a finite density of states even at zero bias, implying that these regions had become metallic after a forming voltage had been applied. At a first glance, the positions of atoms in the modified (red circle) and the pristine (green circles) regions occur identical (C$_3$ symmetry) as shown in the zoom-in Fig. 3(d). However, by comparing the positions of atomic sites along the red and green dashed lines, a clear distortion between tellurium atoms in the pristine and voltage disturbed areas is apparent. The atomic rows of Te in the pristine and voltage modified areas are rotated relative to each other by 3° to 6° as shown in Fig. 3(d) and S4(b). The same phenomenon is observed when Mo$_{0.96}$W$_{0.04}$Te$_2$ devices are analyzed by STM (see supplementary material section S5 for details). Last, it is important to note that – like in the RRAM cell – metallic features that were created by applying a voltage of -3 V can be turned semiconducting again by applying a voltage of opposite polarity. Since the change of surface topography is highly non-uniform however, it is at this stage not possible to unambiguously follow one particular STM feature exclusively through the setting and resetting procedure. The sum of these observations lead us to believe that MoTe$_2$ undergoes a reversible phase transition from a semiconducting to a metallic state under application of an electric field, which is responsible for the RRAM behavior observed by us and reported here.

Next, it is worth asking the question whether any stable metallic phase is expected to exist in MoTe$_2$ that can be created by means of an electric field. Indeed, electrically induced phase transitions were observed previously in TaSe$_2$[21] and TaS$_2$[22]. Moreover, DFT calculations [7] predict that the semiconducting 2H phase in MoTe$_2$ is energetically different by only 31 meV per formula unit from the semi-metallic monoclinic 1T′ phase, a value much smaller than for other TMDs. While these facts speak potentially for an electrically induced 2H-to-1T′ phase transition, one would expect to observe a rectangular surface unit-cell in the areas of the protrusions which is characteristic for the 1T′ phase (see Fig. S7), while our STM image supports more a rotation of the hexagonal 2H-MoTe$_2$ surface cell between the pristine and the voltage modified areas. On the other hand, the 1T′ phase is a modified 2H phase with distinct atomic shifts, and the projected structures, e.g. of Te, on (001) for both 2H and 1T′ have rather similar lattice constants, with a=3.52 Å for the 2H hexagonal phase and a=3.47 Å, b=6.33 Å for the 1T′ pseudo-hexagonal phase. It is thus possible that the topographical changes hinted at above mask the clear distinction between those two phases. An alternative explanation, involves an electrical field induced local rotation of the topmost 2H-MoTe$_2$ planes, resulting in the observed angle between Te-rows in the pristine and modified areas as described above (see Fig. 3(d) and Fig. S4(b)). Indeed, similarly rotated structures have been reported after electron irradiation of 2H-MoTe$_2$[23]. Since the atomic arrangement in the modified and the pristine regions occur similar as shown in the zoom-in Fig. 3(d), we conclude at this stage that a metallic hexagonal 1T phase or a metallic distorted 2H' phase has been created by applying an electric field.

To identify the exact nature of the observed field induced phase change further, scanning transmission electron microscopy (STEM) of cross-sectional samples was utilized for both MoTe$_2$ and Mo$_{1-x}$W$_x$Te$_2$ devices. Before performing the STEM analysis, RRAM devices underwent the same forming process as described above to create filaments in the flakes. Fig. 4(a) shows a cross-sectional atomic-resolution high-angle annular dark field (HAADF)-STEM image of a Mo$_{0.96}$W$_{0.04}$Te$_2$ device. The RRAM multilayer structure is clearly visible from the HAADF contrast (see also Fig. S8-1(b)). Note that the STEM image displays the TMD flake both in the active Ni/Ti/Mo$_{0.96}$W$_{0.04}$Te$_2$/Au/Ti/SiO$_2$ (right) and non-active region (left) where an SiO$_2$ isolation layer on top of the TMD prevented RRAM operation. For all scans performed, we observe that the non-active area *only* exhibited the original 2H phase of Mo$_{0.96}$W$_{0.04}$Te$_2$. On the other hand, in the active region, two structurally distinct domains are observed; the domains are marked as 2H and 2H' in Fig. 4(b), which is a zoom-in HAADF image of the red-marked box in (a). The 2H' region, delineated by the white dash-dotted lines, is about 80 nm wide (consistent with the diameter of the filament measured by C-AFM) and extends vertically through the whole Mo$_{0.96}$W$_{0.04}$Te$_2$ flake. Fig. 4(c) of the 2H region shows an atomic HAADF image in [110]$_{2H}$ zone axis with well-resolved atomic columns of Mo/W and Te. Fig. 4(d) shows a structural HAADF image from the 2H' domain. Instead of the well-aligned atomic columns as observed for the 2H structure, the atomic columns of the 2H' structure are not resolved in this orientation due to apparent shift and splitting along the *c*-direction (The same transition occurs in the vertical MoTe$_2$ device as apparent from Fig. S8-1).

In an attempt to achieve better structural resolution of the 2H' structure, the TEM sample was tilted 30° around the *c*-axis of the 2H to $[120]_{2H}$ zone axis. For the 2H structure, Fig. 4(e) shows again well-resolved atomic columns of the 2H phase. As for the 2H' structure, Fig. 4(f) shows a distinct 'splitting' for both Te and Mo/W atomic columns, which suggests that the atoms in each atomic column experienced a substantial relative displacement, primarily along the *c*-direction, upon electric field application. This finding is consistent with the plain-view STM observation in Fig. 3(d) and Fig. S4(b) shows that the atomic rows of tellurium in the pristine and electric field modified areas are relatively rotated by 3° to 6°. Nano-beam diffraction (NBD) was further performed in an attempt to understand the details of the 2H' structure. Figs. 4(g) and (h) are NBD patterns taken from the 2H' structure with the neighboring 2H structure in $[110]_{2H}$ and $[120]_{2H}$ zone axes respectively. There are no additional reflections beyond the 2H superlattice reflections observed, neither along the rows indicated by yellow arrows nor between the reflections as illustrated by the blue arrows in Fig. 4(h). The absence of such extra-reflections precludes the presence of the well-known 1T, 1T' and Td phases in the region of the 2H' structure (please refer to the structural projections and corresponding electron diffraction patterns in Fig. S8-2 and Fig. S8-3). Although the atomic images of the 2H' structure in Fig. 4(d) and 4(f) show substantial differences from the 2H structure, no apparent differences could be recognized in the NBD patterns taken from the 2H' structure and neighboring 2H structure. These observations suggest that the 2H' phase is a distorted metallic modification of the 2H structure, perhaps some transient state with atoms displaced toward one of the lower symmetry structures, but still within the crystal symmetry of the 2H structure.

It is apparent that the observed phase transition is unique in that it is reversible *and* does not involve a change from an amorphous to a crystalline state as in conventional phase change materials (PCMs)[6]. Instead, $MoTe_2$ and $Mo_{1-x}W_xTe_2$ remain crystalline when undergoing the local semiconductor-to-metal transition from a semiconducting 2H to a metallic 2H′ phase. In comparison to e.g. $VO_2$ that can also undergo an insulator-to-metal phase transition under an electric field[24] but requires a hold current and exhibits a unipolar switching behavior, the $MoTe_2$ phase transition reported here is bipolar, nonvolatile and it occurs at room-temperature.

**Performance of 2H-$MoTe_2$ based RRAM under pulsed operation**
Last, to further substantiate the stable operation in the electric field induced phases of $MoTe_2$, we have performed pulse-measurements, where devices were continuously switched between a high resistive and a low resistive state as shown in Fig. 5(a) by applying a set and reset voltage of 1.7 V and -1.4 V, respectively. Stable operation between multiple pulses is clearly visible with resistance values that are reproducible within around 2 %. Moreover, multilevel stable resistance values can be programmed into the device when higher set or reset voltages (here 1.8 V and -1.7 V) are applied. Fig. 5(b) shows stable resistive states after various short (80 μs) and long (560 μs) voltage pulses that were read at a 0.2 V level. Long pulses result in a more substantial change (training) of the resistive state of the system. This behavior is similar to the potentiation and depression of biological synapses and hints at yet another application space of this class of vertical TMD devices in the realm of neuromorphic computing.

**Summary**

In this paper, an electric field induced reversible semiconductor (2H)-to-metal (2H′) phase transition in vertical 2H-MoTe$_2$ and Mo$_{1-x}$W$_x$Te$_2$ has been achieved experimentally. Conductive filaments were created during electric field application as shown by C-AFM, STM, STS and STEM measurements. Atomic resolution images support that a distorted metallic 2H′ phase is created after voltage modification and that this crystalline phase is responsible for the observed RRAM behavior in MoTe$_2$ and Mo$_{1-x}$W$_x$Te$_2$. Programming voltages are tunable by the MoTe$_2$ flake thickness and partial substitution of Mo by W in MoTe$_2$ reduces the critical electric field for the phase transition in qualitative agreement with DFT calculations. Our work indicates the possibility to locally and selectively engineer the phases in TMDs by electric fields, and demonstrates the potential of TMDs for RRAM applications.

**Materials and Methods[25]**

<u>Devices fabrication and electrical measurement</u>
First a layer of Ti/Au (10 nm/25 nm) which acts as a bottom electrode was deposited onto a 90 nm silicon dioxide (SiO$_2$) layer located on top of a highly doped silicon wafer. Next, TMD flakes from either (i) MoTe$_2$ (2D Semiconductors), (ii) WSe$_2$ (HQ Graphene), (iii) MoS$_2$ (SPI Supplies) or (iv) Mo$_{1-x}$W$_x$Te$_2$ (NIST) were exfoliated onto this electrode using standard scotch tape techniques, followed by a thermal evaporation of 55 nm SiO$_2$ acting as an insulating layer. The device fabrication was finished by the definition of a Ti/Ni (35 nm/50 nm) top electrode. Electrical characterization of the devices was performed at room-temperature using a parameter analyzer (Agilent 4156C).

<u>TMDs synthesis and characterization</u>
Both 1T′- and 2H-Mo$_{1-x}$W$_x$Te$_2$ crystals (x = 0, 0.03, 0.04, 0.07, 0.09) were produced at NIST using the Chemical Vapor Transport (CVT) method. First, polycrystalline Mo$_{1-x}$W$_x$Te$_2$ powders were synthesized by reacting stoichiometric amounts of molybdenum (99.999 %), tungsten (99.9 %) and tellurium (99.9 %) at 750 °C in a vacuum-sealed quartz ampoule. Next, Mo$_{1-x}$W$_x$Te$_2$ crystals were grown at 950 °C to 1000 °C using approximately 1 g of poly-Mo$_{1-x}$W$_x$Te$_2$ charge and a small amount of iodine (99.8 %, 5 mg/cm$^3$) sealed in evacuated quartz ampoules. The ampoules were ice-water quenched after 7 days of growth yielding Mo$_{1-x}$W$_x$Te$_2$ crystals in the metallic 1T′ phase. The 1T′-MoTe$_2$ (Mo$_{1-x}$W$_x$Te$_2$, x ≠ 0) crystals were then converted to the semiconducting 2H phase by annealing in vacuum-sealed ampoules at 950 °C (750 °C) for 24 h (72 h) followed by cooling to room temperature at a 10 °C/h rate. 2H-MoTe$_2$ crystals were also obtained by CVT growth at 800 °C for 140 h using TeCl$_4$ (99.9 %, 5.7 mg/cm$^3$) as a transport agent. At this temperature, MoTe$_2$ crystals grow directly in the 2H phase and hence do not experience a 1T′-2H transition during cooling.

Crystal phases and chemical compositions of $Mo_{1-x}W_xTe_2$ samples were determined by powder X-ray diffraction and energy-dispersive X-ray spectroscopy, respectively. More information on crystal preparation and characterization can be found elsewhere[26].

Conductive AFM measurement

C-AFM was performed in a Veeco Dimension 3100 AFM system. All AFM images were taken in the contact mode using SCM-PIT tips (Bruker). The conductive tip consists of 0.01 - 0.025 Ohm·cm antimony-doped Si coated with PtIr.

STM and STS characterization

2H-$MoTe_2$ and 2H-$Mo_{0.96}W_{0.04}Te_2$ flakes were exfoliated from a 2H-$MoTe_2$ and 2H-$Mo_{0.96}W_{0.04}Te_2$ bulk crystals (2D Semiconductors and NIST respectively) onto Au pads using standard scotch tape techniques. An Omicron ultrahigh-vacuum (UHV) STM was used to perform the surface analysis of $MoTe_2$ and $Mo_{0.96}W_{0.04}Te_2$ at room temperature. During all measurements, the electrochemically etched tungsten tip was grounded and the voltage was applied to the Au pad. All STM images were recorded at a tunneling current of 2 nA and a bias voltage of -0.9 V. The STM data were analyzed with WSxM software[27].

TEM sample preparation with SEM/FIB

FEI Nova NanoLab 600 DualBeam (SEM/FIB) was employed to prepare cross-sectional TEM samples. Carbon was initially deposited on top of the device to protect the samples surface. To reduce Ga-ions damage, in the final step of preparation the TEM samples were thinned with 2 kV Ga-ions using a low beam current of 29 pA and a small incident angle of 3 degree. An FEI Titan 80-300 probe-corrected STEM/TEM microscope operating at 300 keV was employed to acquire both nano-beam diffraction patterns and TEM images in TEM mode as well as atomic-resolution high-angle annular dark field (HAADF) images in STEM mode.

**Acknowledgements**

This work was in part supported by the STARnet center FAME, a Semiconductor Research Corporation program sponsored by MARCO and DARPA. S.K. acknowledges support from the U.S. Department of Commerce, National Institute of Standards and Technology under the financial assistance award 70NANB16H043. H. Z. acknowledges support from the U.S. Department of Commerce, National Institute of Standards and Technology under the financial assistance awards 70NANB15H025 and 70NANB17H249. A. V. D. and S. K. acknowledge the support of Material Genome Initiative funding allocated to NIST. We thank Irina Kalish (NIST) for conducting XRD and EDS on $Mo_{1-x}W_xTe_2$ samples.

**Author contributions**

F. Z. and J. A. designed the experiments. F. Z. fabricated, measured the devices and performed the Conductive AFM measurement. S. K. and A. V. D. synthesized $Mo_{1-x}W_xTe_2$ alloy samples. C. A. M. and D. Y. Z. performed the STM and STS measurements. D. Y. Z. contributed the STM surface analysis. H. Z. prepared TEM samples using SEM/FIB and performed TEM/STEM measurements. H. Z., L. A. B. and A. V. D. performed the TEM/STEM analysis. F. Z. and J. A. wrote the manuscript and discussed the results at all stages.

**SUPPLEMENTARY MATERIALS**
Supplementary Text
Figs. S1 to S8

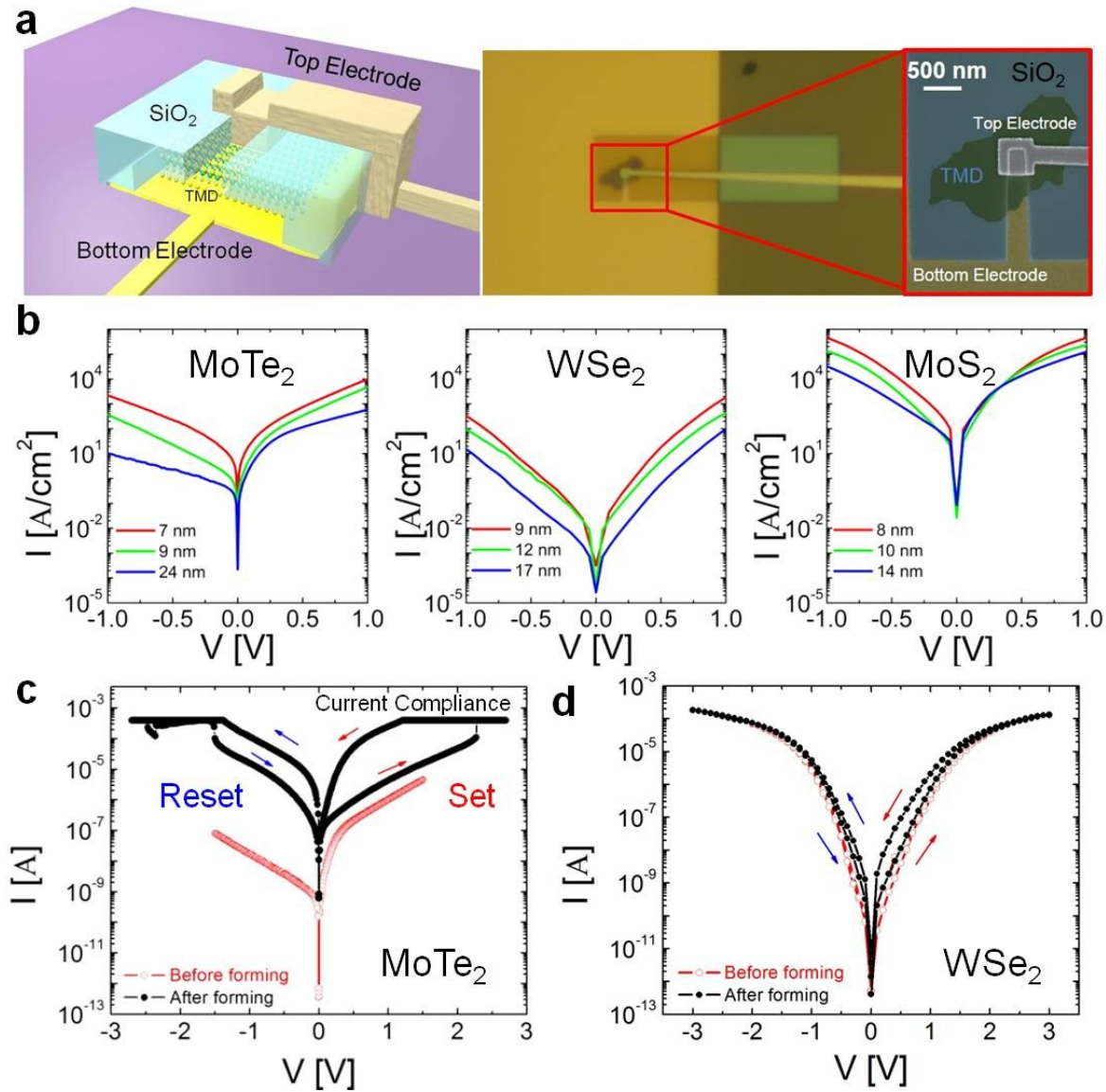

**Fig. 1. Vertical TMD based device characterization.** (**a**) Schematic diagram of a vertical TMD device and optical and SEM images showing the top (Ti/Ni) and bottom (Ti/Au) electrodes, and the SiO$_2$ isolation layer as well as the actual flake. (**b**) Area normalized I-V curves of vertical MoTe$_2$, WSe$_2$, and MoS$_2$ devices before electroforming for different flake thicknesses. (**c**) I-V curves of a vertical MoTe$_2$ device from a flake with a thickness of 24 nm and a contact area of 520 nm x 330 nm. Red circles show I-V curves before memristive switching occurred. The solid black dots show the current versus voltage dependence after forming. Arrows indicate the sweep direction of the applied DC voltage. The current compliance is set to 400 μA. (**d**) Exemplary I-V curves of a vertical WSe$_2$ device with flake thickness of 9 nm. Red circles show I-V curves before RRAM formation and the solid black dots show the current versus voltage dependence after forming.

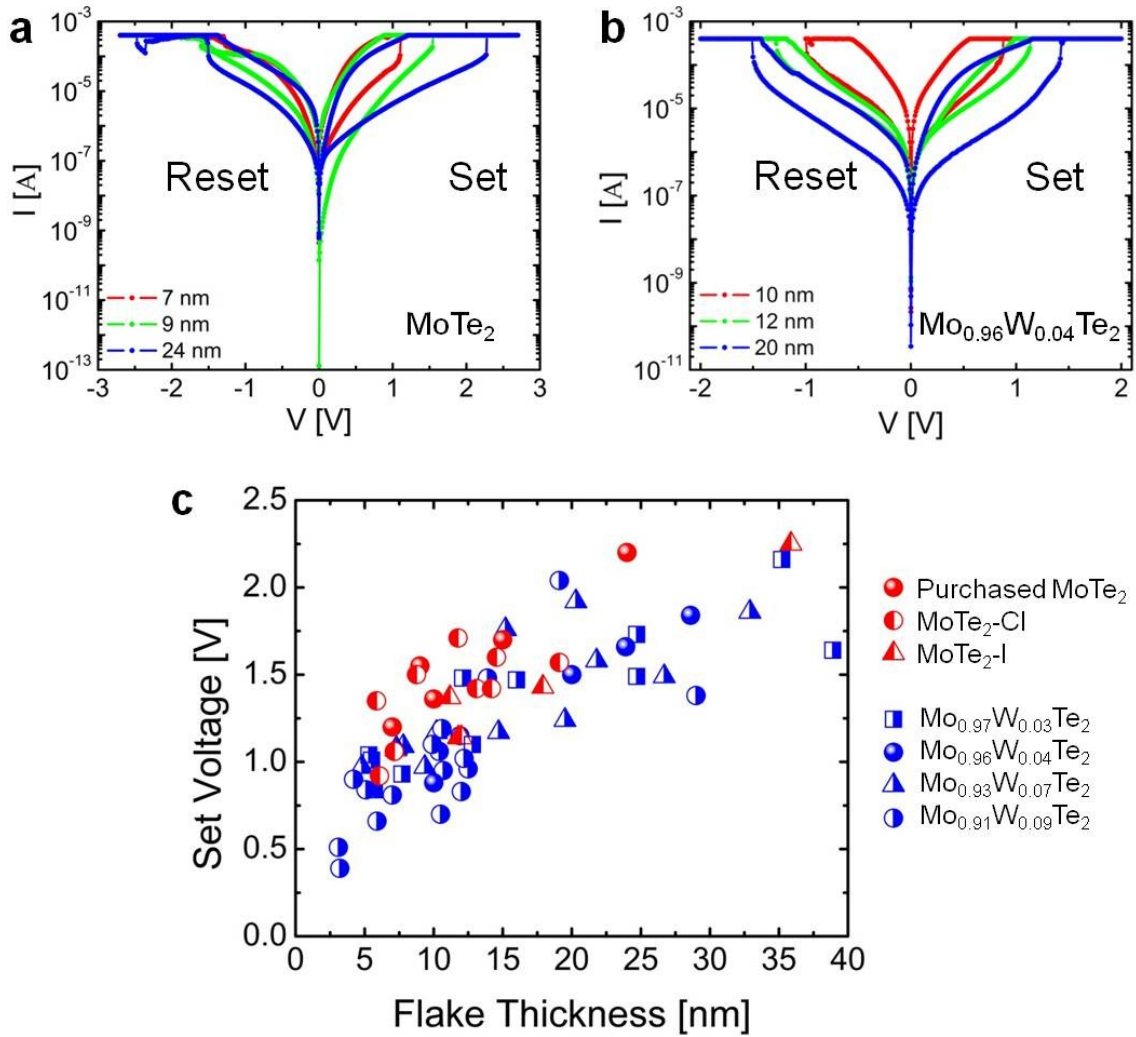

**Fig. 2. 2H-MoTe$_2$ and 2H-Mo$_{1-x}$W$_x$Te$_2$ based RRAM behavior and their set voltages as a function of flake thickness.** (**a**) Log scale I-V curves of vertical MoTe$_2$ RRAM devices after electroforming. The active device area of the 7 nm, 9 nm, and 24 nm MoTe$_2$ flake devices are 542 nm x 360 nm, 542 nm x 360 nm and 518 nm x 332 nm respectively. (**b**) Log scale I-V curves of vertical Mo$_{0.96}$W$_{0.04}$Te$_2$ RRAM devices after electroforming with a current compliance of 400 µA. The active device area of the 10 nm, 12 nm, and 20 nm Mo$_{0.96}$W$_{0.04}$Te$_2$ flake devices are 500 nm x 380 nm, 522 nm x 400 nm and 510 nm x 384 nm respectively. (**c**) Set voltage values scale with the flake thickness of MoTe$_2$, Mo$_{0.97}$W$_{0.03}$Te$_2$, Mo$_{0.96}$W$_{0.04}$Te$_2$, Mo$_{0.93}$W$_{0.07}$Te$_2$ and Mo$_{0.91}$W$_{0.09}$Te$_2$. The error bars of the set voltages and the flake thicknesses are in the range of the dots' sizes. MoTe$_2$-Cl and MoTe$_2$-I denote crystals grown with TeCl$_4$ and I$_2$ transport agents, respectively.

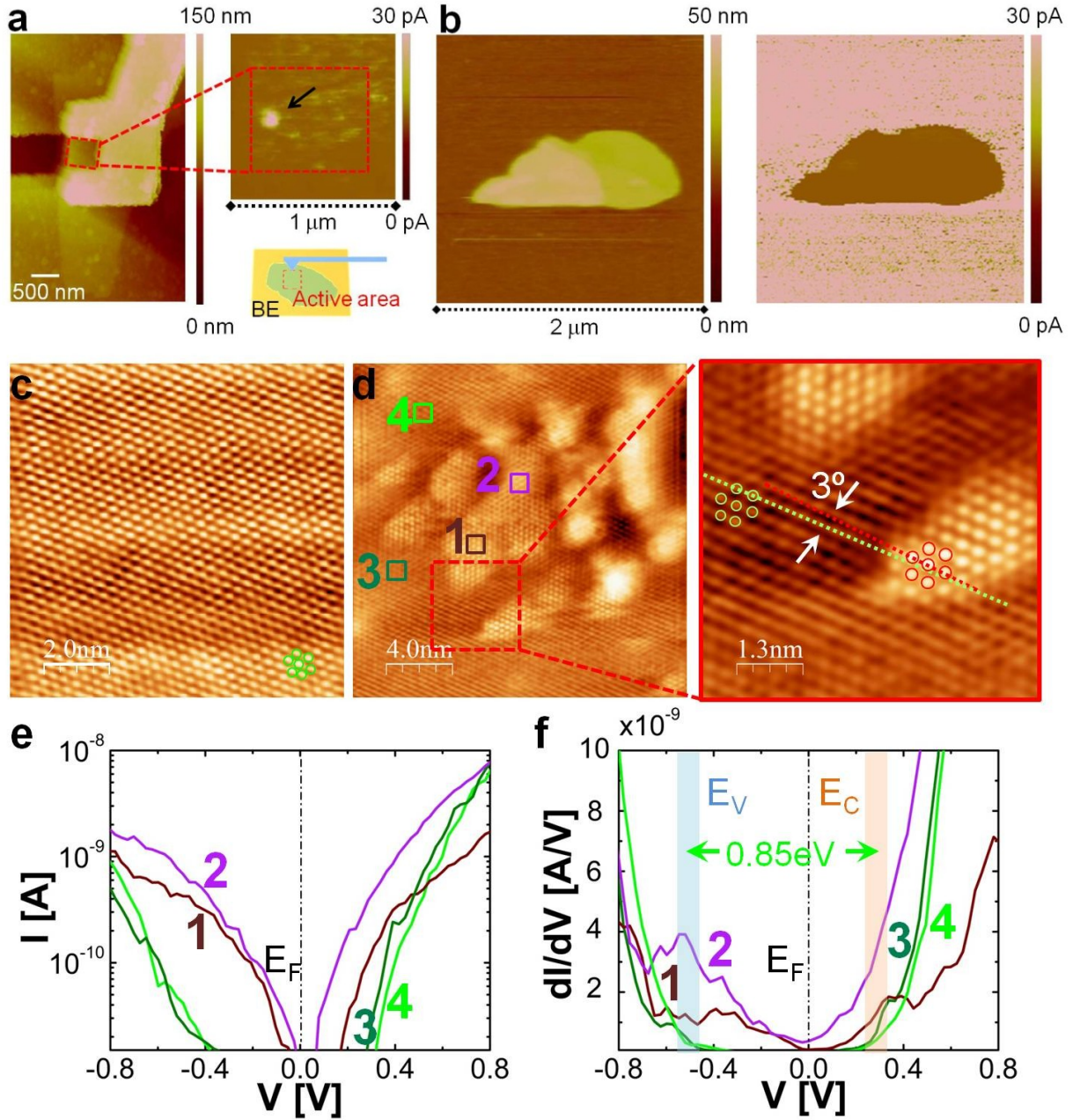

**Fig. 3. C-AFM, STM and STS measurements.** (**a**) Current mapping of a MoTe$_2$ flake after the set process and the formation of the LRS has occurred using a conductive AFM (C-AFM). The red dashed square denotes the active device area before removal of the top electrode. Note the bright spot marked with an arrow that we interpret as a filament. (**b**) C-AFM images of a pristine MoTe$_2$ flake (left: topography and right: current map) showing no indication of the aforementioned highly conductive area. (**c**) STM image (filtered) of the pristine MoTe$_2$ surface. (**d**) STM image (filtered) of a portion of the surface region in (**c**) after a voltage of -3 V was applied to a contact underneath the TMD relative to the STM tip. The zoom-in image indicates that the position of Te atoms has changed after voltage application. The Te rows in the voltage modified region are rotated ~3° relative to the atomic rows in the pristine part. However, the C$_3$

symmetry of the atomic lattice is still clearly visible (green circles for pristine part and red circles for modified area). All images were recorded at tunneling currents of 2 nA and a bias voltage of -0.9 V. (e) shows I-V characteristics obtained by STS measurements corresponding to locations 1 through 4 in (d). (f) shows the corresponding dI/dV spectra with the blue band indicating the position of the valance band edge and the orange band indicating the position of the conduction band edge for the pristine MoTe$_2$ in agreement with the I-V characteristics of locations 3 and 4. Note that 1 and 2 clearly show the absence of a bandgap after voltage application.

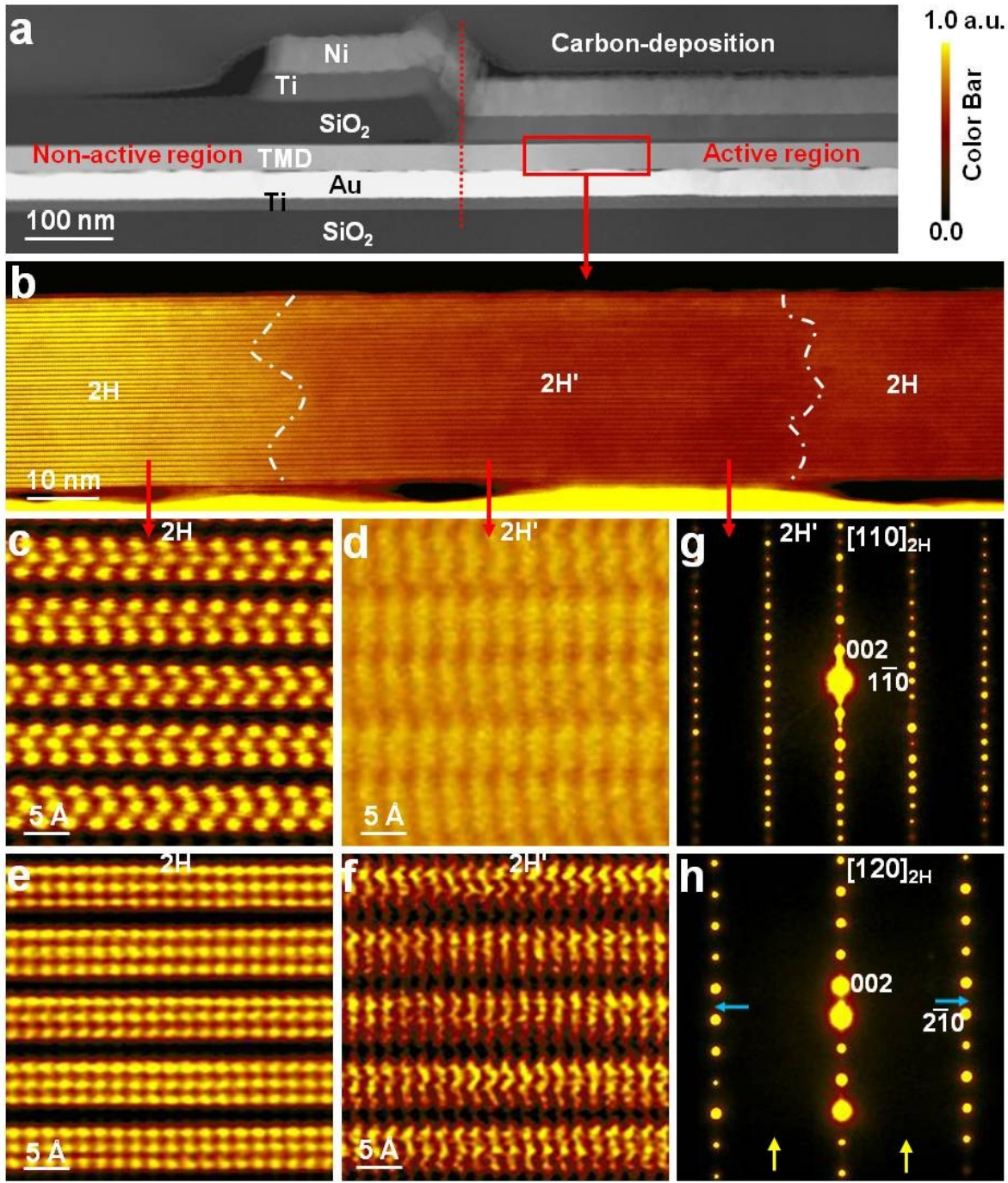

**Fig. 4. STEM measurement and analysis.** (**a**) HAADF-STEM image showing cross-section of the $Mo_{0.96}W_{0.04}Te_2$ device. (**b**) Higher magnification HAADF image from the region defined by a red box in (**a**) and showing co-existence of a distorted structure (2H') with 2H. (**c, d, e, f**) Atomic-resolution HAADF images taken along the $[110]_{2H}$ zone-axis (**c, d**) and $[120]_{2H}$ zone-axis (**e, f**), showing the intact 2H and distorted 2H' structures respectively. (**g, h**) Corresponding nano-beam diffraction pattern taken from the distorted 2H' area, which is still indexed as the 2H structure. False colors are added to aid the eye.

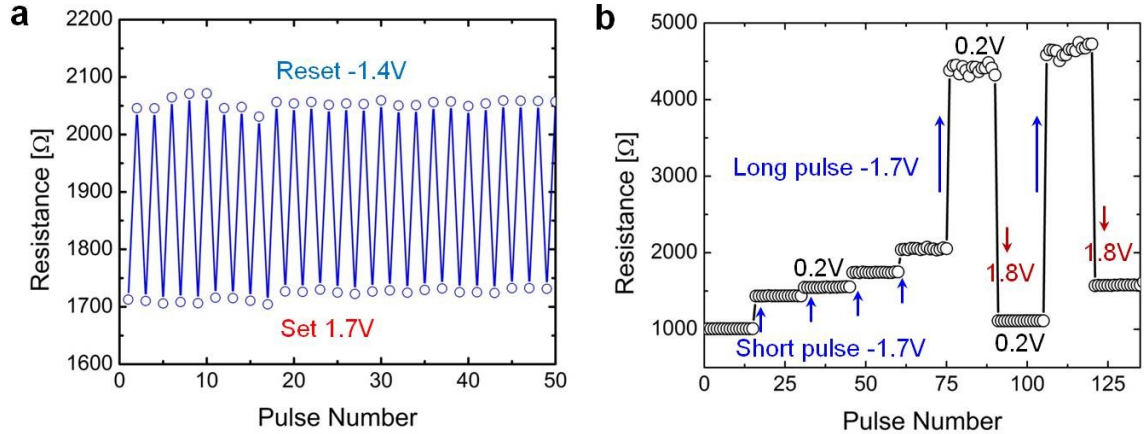

**Fig. 5. Performance of 2H-MoTe$_2$ based RRAM under pulsed operation.** (**a**) Pulse switching performance of a vertical MoTe$_2$ device with a flake thickness of 15nm. The device can be repeatedly switched between a HRS and LRS with 80 μs long voltage pulses. The set voltage of this device is 1.7 V, and the reset voltage is -1.4 V. Read out occurs with the respective set and reset voltage values. (**b**) shows multiple stable states of a device under various set and reset voltage pulses have been applied. read out occurs at a voltage of 0.2 V. Every state is characterized by 15 subsequent read outs. A short pulse (80 μs) of a reset voltage of -1.7 V switches the device into a higher resistance state and longer pulses (560 μs) result in larger changes of the resistance, indicating that the duration of "training" affects the device state. A short pulse at a set voltage of +1.8 V can tune the resistance back to a low resistive state.

# Supplementary Materials for

**Electric field induced semiconductor-to-metal phase transition in vertical MoTe$_2$ and Mo$_{1-x}$W$_x$Te$_2$ devices**

Feng Zhang[1,2], Sergiy Krylyuk[4,5], Huairuo Zhang[4,5*], Cory A. Milligan[1,3], Dmitry Y. Zemlyanov[1], Leonid A. Bendersky[5], Albert V. Davydov[5] and Joerg Appenzeller[1,2*]

*correspondence to: appenzeller@purdue.edu (J.A.); huairuo.zhang@nist.gov (H.Z.)

**This PDF file includes:**

Supplementary Text

Figs. S1 to S8

**Supplementary Text**

Section S1: MoTe$_2$ based RRAM electroforming process

Pristine metal-MoTe$_2$-metal devices exhibit reproducible I-V curves shown as pink dotted line in Fig. S1(a) and (b) as long as a critical, TMD thickness dependent forming voltage is not reached. Measuring beyond this forming voltage results in the RRAM behavior as described in the main text, triggering the resistive switching behavior. External electric fields and Joule heating are both important factors during the formation of conductive filaments. Once a filament is formed, a set voltage lower than the forming voltage is used to switch between the HRS and LRS of the cell. When a reverse polarity electric field is applied, rupture of filaments causes the back transition to the HRS. Note that the set voltage has a linear dependence on the flake thickness (see Fig. 2(c)) indicating that a critical electric field is needed to trigger the memristive behavior. Cycling the MoTe$_2$-based RRAM cells multiple times as shown in Fig. S1(b) results in similar and stable performance specs with set voltages varying by about 0.2 V.

Section S2: Current compliance setting effect on RRAM performance

For a metal oxide based RRAM cell, the LRS resistance can be controlled by the set current compliance which in turn determines the diameter and/or the number of conductive filaments formed, and the HRS resistance can be changed by the reset voltage through the modulation of the ruptured filament length. MoTe$_2$-based RRAM cells, as presented here, show the same behavior as evident from Fig. S2(a). As expected, the higher the current compliance, the lower the LRS resistance becomes. The flake thickness for the device in Fig. S2 is 7 nm. Fig. S2(b) shows the performance of the same device under a 0.5 V read disturb with the current compliance set to 800 μA. Both states show a stable resistance value over 1000 s at room temperature.

Section S3: Performance of 2H-MoTe2 based RRAM under pulsed operation

As discussed in the main text, 2H-MoTe$_2$ based RRAM cells exhibit stable performance specs under various voltage pulse conditions. A reproducible set and reset of the device between a low resistance state and high resistance state is achievable when short voltage pulses are used to manipulate the RRAM state. Moreover, another "pair" of stable resistance values can be programmed into the device when higher set voltages (here 1.8 V) are applied. Fig. S3 shows how pulses of 1.8 V can adjust the device to operate at multiple resistance pairs (HRS/LRS) in a well-controlled tunable fashion without loss of resistance reproducibility. Note that different from the example in the main text, read out of the resistive state of the RRAM cell occurred using the same high set and reset voltages (+1.7 V and -1.4 V respectively) as used for the set and reset procedure itself.

Section S4: MoTe$_2$ switching mechanism exploration by STM

2H-MoTe$_2$ flakes were exfoliated from a 2H-MoTe$_2$ bulk crystal (2D Semiconductors) onto Au pads using standard scotch tape techniques. An Omicron LT ultrahigh-vacuum (UHV) STM was used to perform a surface analysis of the MoTe$_2$ structures at room temperature. During all measurements, the homemade tungsten tip is grounded, and the voltage is applied to the Au pad. All STM images were recorded at a tunneling current of 2 nA and a bias voltage of -0.9 V. Multiple scans were performed on pristine surfaces with no changes observed at the scan voltage of -0.9 V. However, when the local electric field exceeded a threshold value, bright spots appeared in the STM images.

When the pristine 2H-MoTe$_2$ layer was scanned after applying -3 V to induce the filament formation, some bright spots were created. A corrugation profile measured along the black dashed line in Fig. S4(a) is shown in Fig. S4(c). 'Protrusions' with a height of 0.3 nm to 0.6 nm on the sample surface are clearly observed. At this stage we cannot distinguish whether the protrusions are a result of topographic changes or related to the LDOS differences between the bright spots and the pristine area. The lattice configuration in the bright spot areas still occurs hexagonal, which is consistent with the symmetry of 2H-MoTe$_2$. However, a rotation seems to have occurred after the voltage modification. From Fig. S4(b), the Te atomic rows marked as red lines in the bright spots are rotated by about 3° to 6° relative to the Te atomic rows (green lines) of pristine parts.

Section S5: Mo$_{0.96}$W$_{0.04}$Te$_2$ switching mechanism exploration by STM

Fig. S5 shows STM and STS results on Mo$_{0.96}$W$_{0.04}$Te$_2$. After -3V voltage modification, bright spots were created and Te atomic rows marked as red lines in the bright area are rotated relative to the Te atomic rows (green lines) of pristine parts as shown in the zoom-in of the Fig. S5(a). Note that this effect is similar to our observations in case of MoTe$_2$. Fig. S5(a) indicates that the pristine regions show the expected C$_3$ symmetry of 2H-Mo$_{0.96}$W$_{0.04}$Te$_2$. Fig. S5(b) is the 3D topography image of S5(a). As we discuss in the main text, we cannot determine at this point in time whether the bright spots are real protrusions or caused by an increase in LDOS or both. Fig. S5(c) shows STS measurements in the pristine area (black) and in the area of a bright spot (red). As in the case of MoTe$_2$, the bright spot area exhibits metallic behavior after voltage modification, which indicates again an electric field induced semiconductor-to-(semi)metal transition.

Section S6: Electrical characterization of the Mo$_{0.96}$W$_{0.04}$Te$_2$ alloy

Fig. S6 displays a typical transfer characteristic of a 2H-Mo$_{0.96}$W$_{0.04}$Te$_2$ field effect transistor (FET), indicating that the ternary compound shows clear semiconducting behavior with device characteristics similar to the case of MoTe$_2$.

Section S7: Structural characterization of 1T′-MoTe$_2$

Fig. S7 displays an STM image of 1T′-MoTe$_2$ showing the expected rectangular surface unit-cell of this particular phase.

Section S8: STEM analysis

The same 2H to 2H′ phase transition discussed in the context of figure 4 in the main text was also observed in vertical MoTe$_2$ devices. Fig. S8-1(a) shows a low-magnification bright-field transmission electron microscopy (TEM) image of an MoTe$_2$ device cross-section. In the image two regions can be identified: Area 1, which corresponds to the original structure of MoTe$_2$ outside of the active electrical contact, and Area 2, which corresponds to the electrically cycled part of the device. Fig. S8-1(b) shows a higher magnification TEM image of the Ni/Ti/MoTe$_2$/Au/Ti/SiO$_2$ multilayered architecture of the active region. The contact between Au and MoTe$_2$ is not continuous and shows some 5 nm to 20 nm gaps distributed along the interface. Fig. S8-1(c) shows a typical nano-beam diffraction (NBD) pattern of the MoTe$_2$ crystal in the Area 2. The pattern can be indexed as the [110] zone-axis of the 2H structure. No apparent differences could be recognized in all the areas by NBD mapping. Atomic resolution HAADF scanning transmission electron microscopy (STEM) was further employed to study the local defects of the MoTe$_2$ crystal in Area 2. Fig. S8-1(d) shows a typical atomic image of the unperturbed MoTe$_2$ crystal in most of the region in Area 2, representing a characteristic atomic structure of [110] zone-axis of the 2H structure (see Fig. S8-2 for structural projections at [110]). In the bottom of the MoTe$_2$ layer, Te atoms connect closely with the Au atoms, as can be seen at the clean atomically-resolved interface between Au and MoTe$_2$, ensuring low resistance contact formation for electrical measurements. Although no difference has been noticed by NBD mapping, a distorted region around 80 nm in width was recognized in the MoTe$_2$ layer stack using atomic resolved HAADF-STEM analysis. Fig. S8-1(e) and S8-1(f) show high magnification atomic images of the distorted structure. As shown by the black circles, all of the original single Te atomic layer shows a double layer structure and a relative shift of the MoTe$_2$ subunit cells can be identified by the round dot and square dot black lines, suggesting that the original atomic sites of the 2H structure were "shuffled" along the c-direction, resulting an obscure gap between the Te-Mo-Te layers. It is difficult to understand precisely the nature of disorder from the STEM characterization. However, the NBD mapping suggests that the 2H′ structure is a distorted modification of the 2H structure, and may be interpreted as a transient state toward the 1T' phase, with changes from semiconducting to metallic behavior.

Figs. S8-2 and S8-3 show the structural projections and corresponding electron diffraction patterns of the 2H and 1T′ variants along <100> and <1-10> zone axes, respectively. By comparing electron diffraction patterns with those shown in Figs. 4(g) and 4(h), we verified that the observed structure is not 2H nor the well-known 1T', 1T or Td phase but instead a new unknown 2H′ phase.

Section S9: C-AFM exploration of the switching mechanism in MoTe$_2$ RRAM

Since the top electrode in our RRAM design prevents a locally resolved scan of the currents through a $MoTe_2$ flake after formation when employing a C-AFM (Veeco Dimension 3100 AFM) measurement, the top electrode had to be carefully removed prior to any further characterization. To do so, we used PMMA as etch mask to define the etching area and Nickel etchant TFB to etch away the top Ni contact and BOE to remove the Ti layer to access the TMD. We carefully checked that this etching approach did not harm the TMD and did not result in any additional features during the C-AFM measurements that would be unrelated to the forming process.

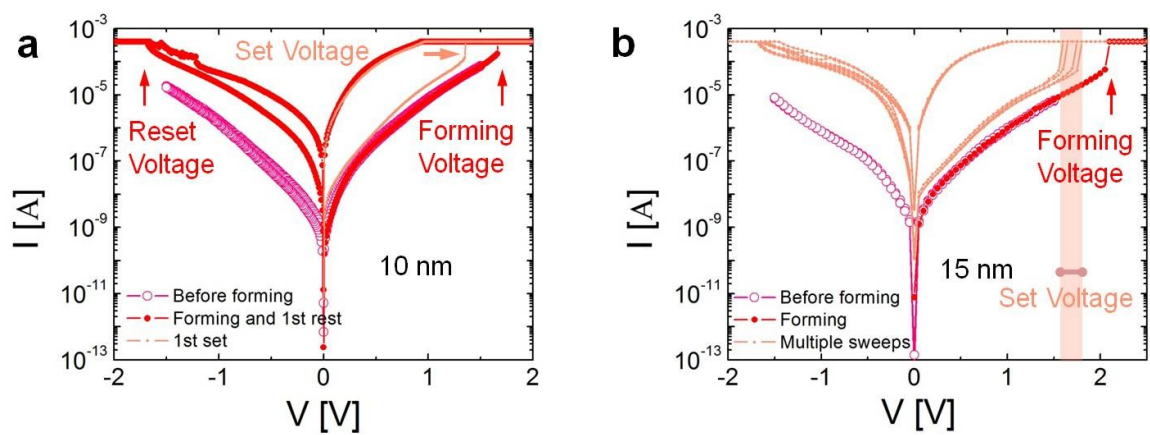

**Fig. S1.** Typical I-V curves of a vertical MoTe$_2$ RRAM showing the forming process. (**a**) RRAM forming for a 10 nm thick flake including first set and reset. (**b**) Multiple sweeps after forming had occurred for a 15 nm flake. Note the rather narrow "band" of set voltages. Current compliance is set to 400 µA for both cases.

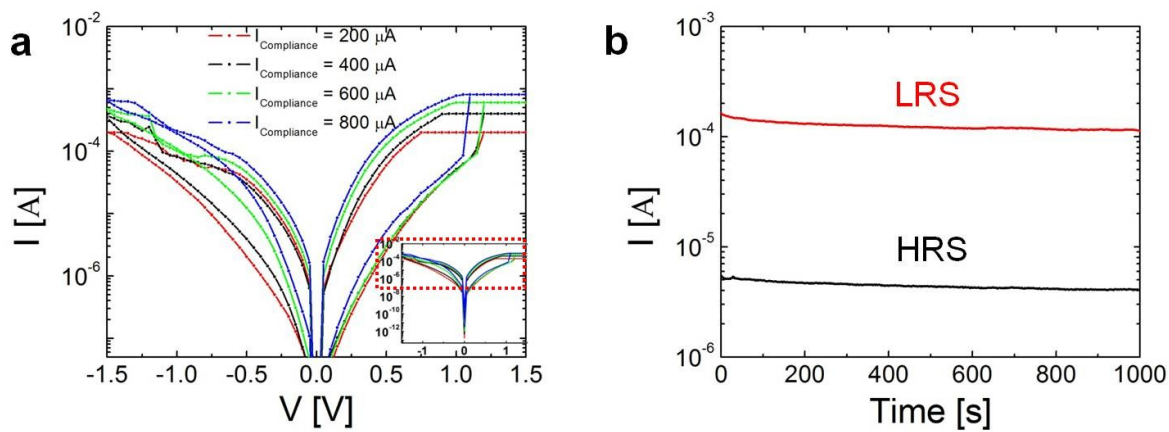

**Fig. S2.** (**a**) DC switching characteristic for a 7 nm thick MoTe$_2$ flake for different current compliances. (**b**) shows read disturb measurements for the same RRAM device at 0.5 V with the current compliance set to 800 µA.

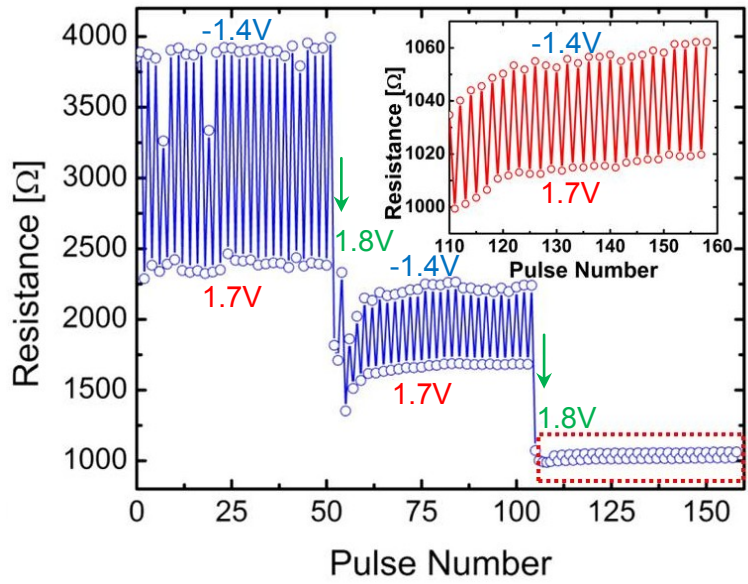

**Fig. S3.** Multilevel characteristics of a vertical MoTe$_2$ device. When a higher set voltage of 1.8 V is applied, the device operates in a new (lower resistive) state that is again stable upon applying set/reset voltages of 1.7 V and -1.4 V respectively. A third state can be "dialed in" through yet another 1.8 V pulse. The inset figure is the zoom-in of the red dashed part.

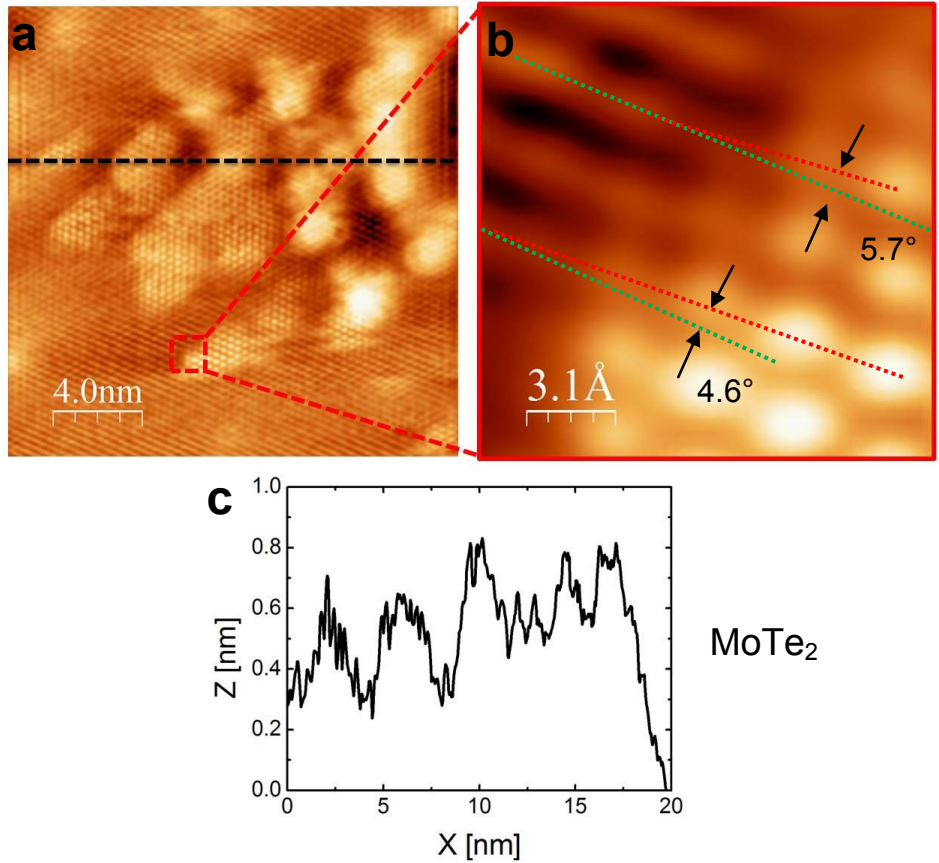

**Fig. S4.** (**a**) STM image (FFT filtered) of MoTe$_2$ after a voltage of -3 V was applied to a contact underneath the TMD relative to the STM tip. Bright spots are created. (**b**) is the zoom-in image of (**a**), indicating that the positions of Te atoms have changed after voltage application. The red lines were drawn along the Te rows after voltage modification and the green lines follow the Te atomic rows of pristine areas. The relative angles between Te-rows are shown in the image. Both images were recorded at tunneling currents of 2 nA and a bias voltage of -0.9 V. (**c**) Corrugation profile along the black dashed line in (**a**).

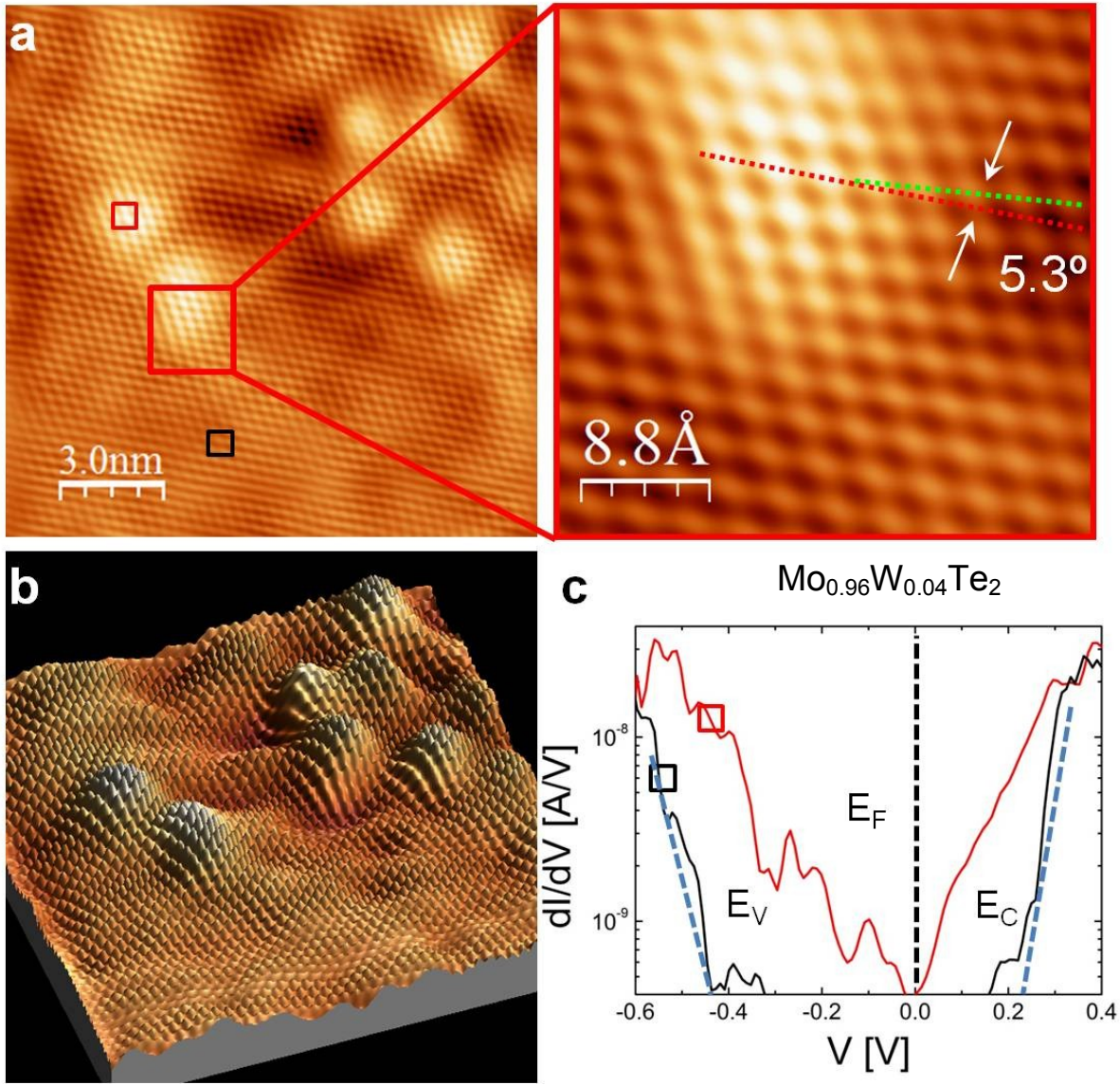

**Fig. S5.** (**a**) STM image (FFT filtered) of $Mo_{0.96}W_{0.04}Te_2$ after a voltage of -3 V was applied to a contact underneath the TMD relative to the STM tip. Bright spots are created, similar to the case of $MoTe_2$ after electric field modification. The zoom-in image indicates that the positions of Te atoms have changed after voltage application. Both images were recorded at tunneling currents of 2 nA and at a bias voltage of -0.9 V. (**b**) is the 3D image of (**a**), if treating the bright areas as protrusions. (**c**) dI/dV spectra corresponding to the marked positions in (**a**). The bright area shows metallic behavior after voltage modification.

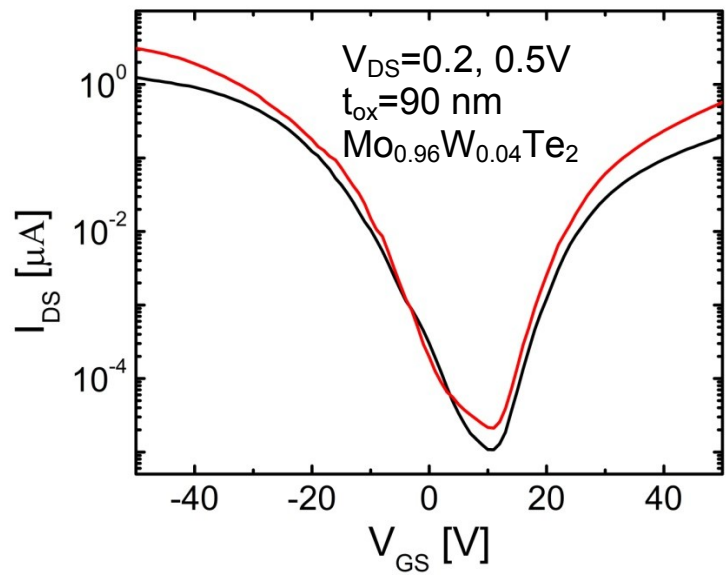

**Fig. S6.** Transfer characteristic of a $Mo_{0.96}W_{0.04}Te_2$ back-gated device.

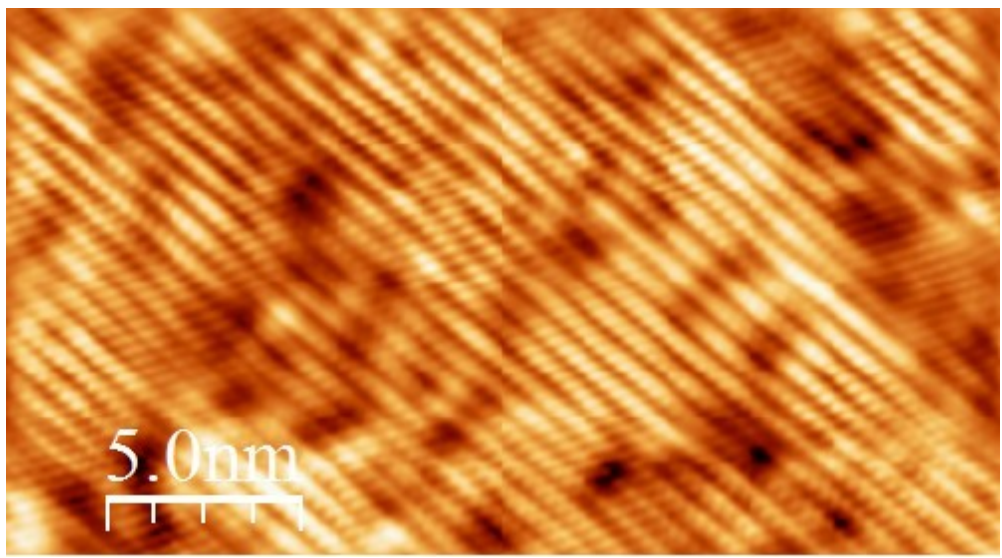

**Fig. S7.** STM image (FFT filtered) of 1T′-$MoTe_2$.

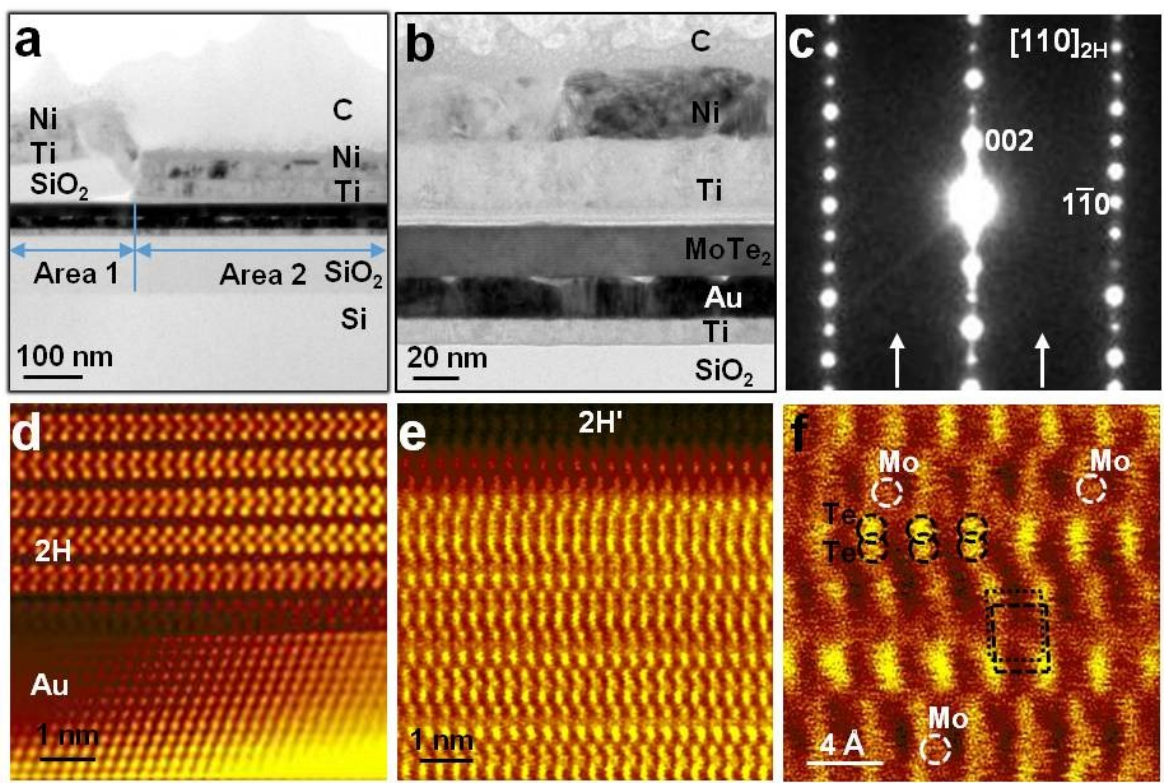

**Fig. S8-1.** (a) Bright-field TEM image showing the cross-section of a MoTe$_2$ device with 'Area 1' denoting the electrically non-active region and 'Area 2' denoting the electrically cycled active region. (b-f) images were taken from Area 2. (b) Enlarged TEM image showing the multilayer architecture of the device. (c) Nano-beam diffraction pattern taken along the [110]$_{2H}$ zone-axis. (d) Atomic resolution HAADF image showing the typical undisturbed 2H structure of the MoTe$_2$ crystal along the [110]$_{2H}$ zone-axis, also showing the atomic contact at the interface of MoTe$_2$ and the bottom Au-electrode. (e, f) HAADF images showing the details of the distorted 2H′ structure. Black and white circles denote the recognizable Te-sites and Mo-sites respectively. Black round dot and square dot lines illustrating the atomic shuffle in the neighboring layers along the view direction.

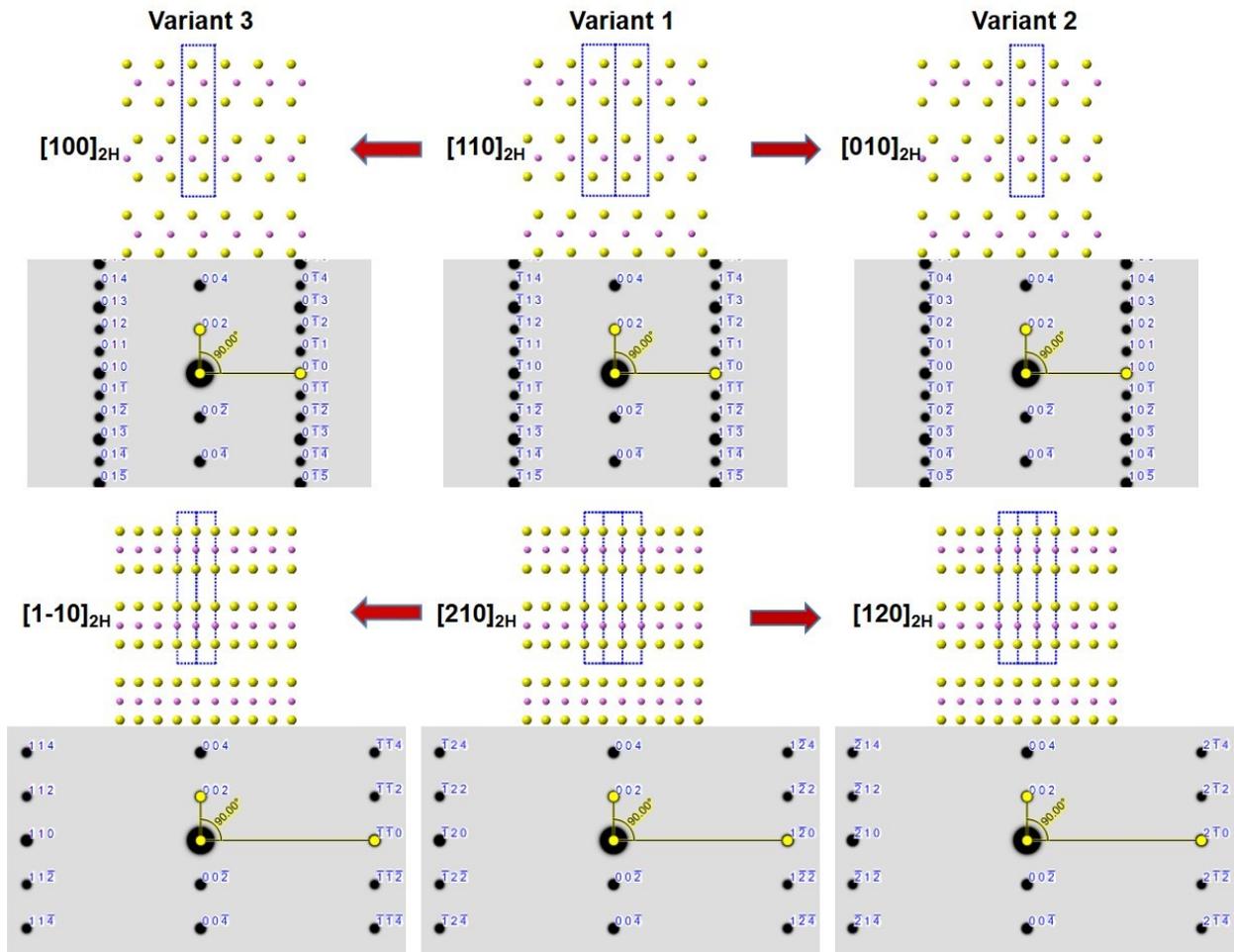

**Fig. S8-2.** Structural projections and corresponding electron diffraction patterns of the 2H variants along <100> and <1-10> zone axes.

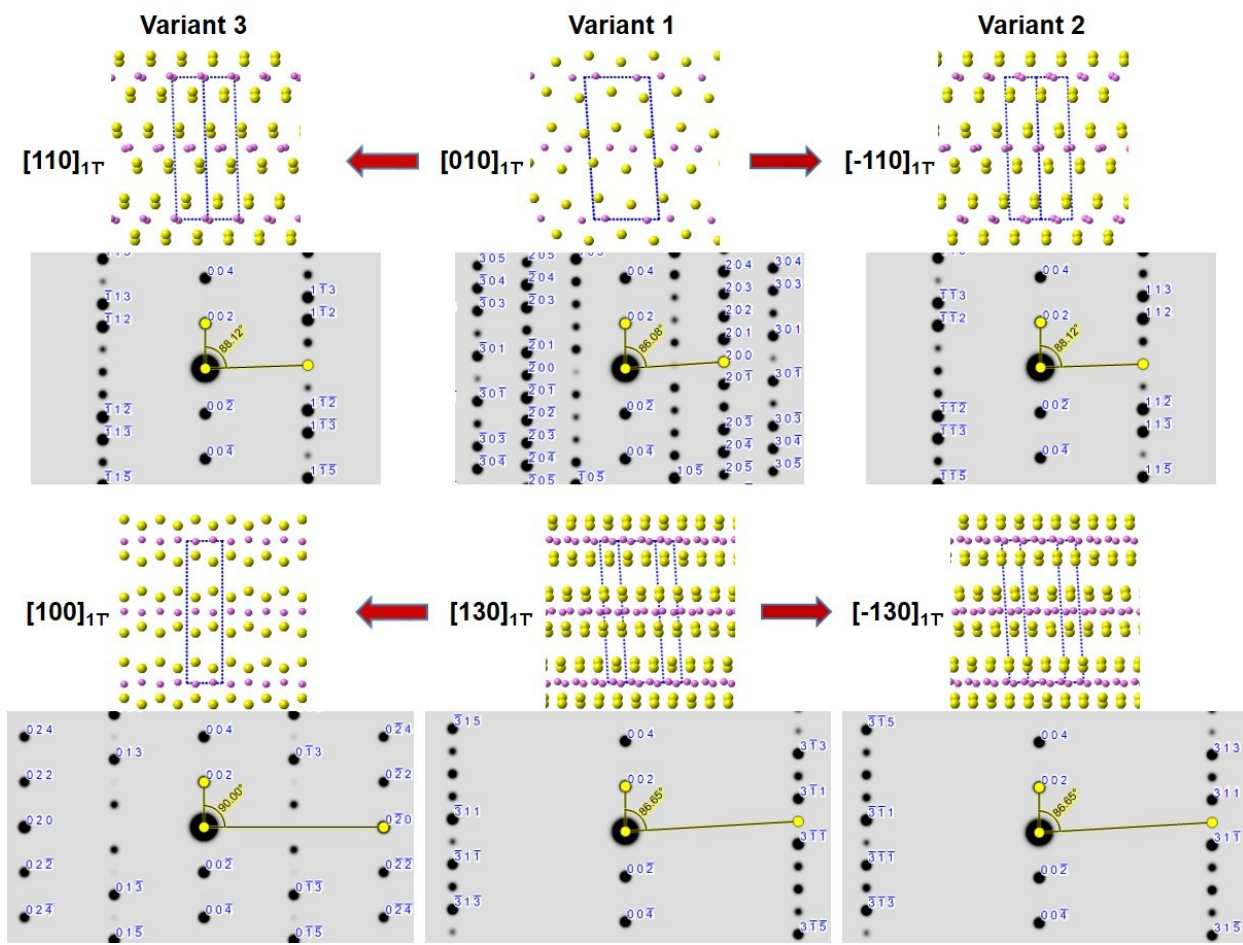

**Fig. S8-3.** Structural projections and corresponding electron diffraction patterns of the 1T' variants in <110> and <100> zone-axes. With a 2H-to-1T' phase transition, 2H <100> variants can be transformed to 1T' <110> variants, and 2H <1-10> variants to 1T' <100> variants.